# CAN ONLINE CUSTOMER REVIEWS HELP DESIGN MORE SUSTAINABLE PRODUCTS? A PRELIMINARY STUDY ON AMAZON CLIMATE PLEDGE FRIENDLY PRODUCTS

**Michael Saidani[1], Harrison Kim**
Department of Industrial and Enterprise
Systems Engineering, University of Illinois at
Urbana-Champaign, Illinois, USA

**Nawres Ayadhi, Bernard Yannou**
Laboratoire Genie Industriel, CentraleSupélec,
Université Paris Saclay,
Gif-sur-Yvette, France

## ABSTRACT

*Online product reviews are a valuable resource for product developers to improve the design of their products. Yet, the potential value of customer feedback to improve the sustainability performance of products is still to be exploited. The present paper investigates and analyzes Amazon product reviews to bring new light on the following question: "What sustainable design insights can be identified or interpreted from online product reviews?". To do so, the top 100 reviews, evenly distributed by star ratings, for three product categories (laptop, printer, cable) are collected, manually annotated, analyzed and interpreted. For each product category, the reviews of two similar products (one with environmental certification and one standard version) are compared and combined to come up with sustainable design solutions. In all, for the six products considered, between 12% and 20% of the reviews mentioned directly or indirectly aspects or attributes that could be exploited to improve the design of these products from a sustainability perspective. Concrete examples of sustainable design leads that could be elicited from product reviews are given and discussed. As such, this contribution provides a baseline for future work willing to automate this process to gain further insights from online product reviews. Notably, the deployment of machine learning tools and the use of natural language processing techniques to do so are discussed as promising lines for future research.*

Keywords: sustainable design, data-driven design, online reviews, customer feedback, ecolabels, sustainability-related comments.

## 1. INTRODUCTION

### 1.1 Motivation and related work

Over the past two decades, with the rapid growth of online shopping, electronic word-of-mouth communication has become an important source of information for consumers planning to purchase new products [1]. Online customer reviews are a form of customer feedback on electronic commerce and online shopping sites. The massive amount of online review data posted every day is a convenient way not only for customers to make their voice heard but it also represents an opportunity for designers to improve the features of the products they create, develop, and put on the market [2]. According to recent surveys conducted in 2020 [1], on the one hand, 41% of all consumers are value-driven consumers who want products and services to simplify their life and are willing to pay for those benefits. This group is not particularly inclined to switch habits to reduce negative environmental impact. On the other hand, 40% of all consumers are purpose-driven consumers who seek products that align with their lifestyle and particularly value the health and wellness benefits, as well as the sustainability-related features offered by certain products.

While the global awareness on environmental issues and the concept of sustainable products (e.g., through ecolabels [3]) are growing [4], the consumer feedback on sustainability-related aspects embedded in products remains to be investigated. Particularly, it could be relevant to question and illustrate how sustainability-related reviews could be used to support the generation of sustainable design leads, e.g., for a new generation of environmentally-friendly products valued by customers. Interestingly, unlike traditional approaches to understand customer choices and perceptions of products – e.g., surveys, focus groups, or interviews – which are often time-consuming, costly, and thus limited in terms of content, online reviews are one of the largest and most accessible collections of crowdsourced customer opinions [2]. The democratization (including the availability, accessibility, and interpretation) of data will become a pillar for new product development processes [5].

---

[1] Contact author: msaidani@illinois.edu

Therefore, online product reviews can be considered a commendable alternative source not only to understand customer needs [6] but also to help designers propose relevant design innovations [7]. Data analytics can reveal usage patterns and drive adaptations of products as well as provide valuable feedback to product developers [5]. For instance, Hou et al. (2019) summarized how online review analysis could be helpful during the product design and development phases [7], including: design trends monitoring, building market strategies, product longevity prediction, construction of product improvement strategy, discovering product defects. Yet, as of now, there is a lack of study trying to link product sustainability features and online review analysis, notably in terms of sustainable design implication [2].

El Dehaibi et al. (2019) started to manually identify and extract customer Perceptions of product Sustainability Features (PerSFs) from online reviews, and modeled these perceptions of product sustainability using machine learning techniques (natural language processing algorithms) to determine which of these features are associated with positive and negative sentiments [2]. Through a single case study on press coffee carafe reviews, their work focused on the three pillars of sustainability. Regarding environmental aspects, customers associated "stainless steel" and "strong glass" for the coffee carafe with positive PerSFs and the "use of plastic" or "product breaking" with negative PerSFs. Note that features typically associated with "real" environmental benefits, such as energy use and water use, were identified but underrepresented as compared to "perceived" features that are not necessarily beneficial to the environment.

While the authors mentioned that "online reviews can enable designers to extract PerSFs for further design study and to create products that resonate with customers' sustainable values" [2], further works are recommended to investigate how such reviews analyses can "feed into design methods" to support and guide the design of more sustainable products.

### 1.2 Research question and objective

In all, while online product reviews can offer a lot of valuable information for designers, translating such pieces of information into (sustainable) design insights is not a straightforward process. Moreover, data is often collected opportunistically according to what can be captured, rather than driven by an understanding of what ought to be captured [5]. A key challenge addressed in this paper is thus to determine what kinds of sustainable design leads can be identified and/or interpreted from customer reviews.

Based on the limitations aforementioned, the overall objective of this project is to take the extraction of customer perceptions of product sustainability from online reviews [2] to the next level, and fill two key research gaps: (i) analyze online product reviews to generate sustainable design leads, and (ii) build a machine learning model to automate this process. The present paper addresses this first research gap, while a complementary paper submitted to this conference discusses the second one [8].

In this line, the research question driving the present research paper is as follows: "What sustainable design insights can be identified or interpreted from online product reviews?". Note that the umbrella expression "sustainable design insights" encompasses here any information that could be used to design more sustainable products throughout their life cycles [9-12]. As illustrated in Figure 1, reviews from Amazon Climate Pledge Friendly products [13] are used to figure out to what extent we can learn from customers' product reviews to design more sustainable products.

The empirical research approach, as well as the resources used to provide new findings on this research question, are detailed in Section 2. Illustrative and concrete examples of sustainable design insights that could be elicited from product reviews are given and discussed in Section 3. Finally, based on the results of this preliminary study, practical implications and promising perspectives for the future of design for sustainability are provided in Section 4.

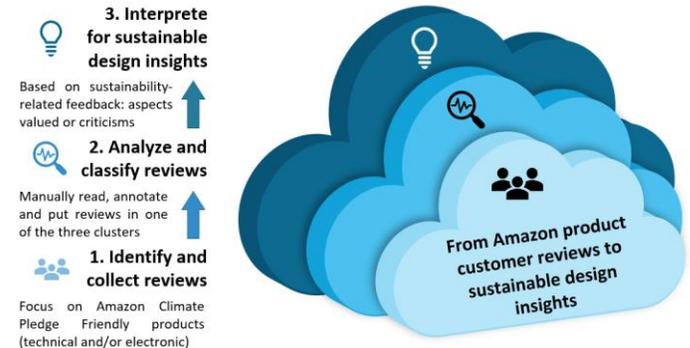

**FIGURE 1:** OBJECTIVE AND WORKFLOW OF THE PAPER

In another paper submitted to this conference [8], a review and critical analysis is conducted on how existing sentiment analysis and opinion mining approaches can be deployed to automatically assess the perception and value of product sustainable features from online customer reviews. It particularly discusses the opportunities and challenges offered of natural language process techniques to automate the possibility of getting sustainable design insights from online product reviews. Note that although these two papers are complementary, they each have their own research findings and contributions, and can therefore be considered stand-alone papers.

## 2. MATERIALS AND METHODS
### 2.1 Data-based research approach
The overall research method is depicted in Figure 2.

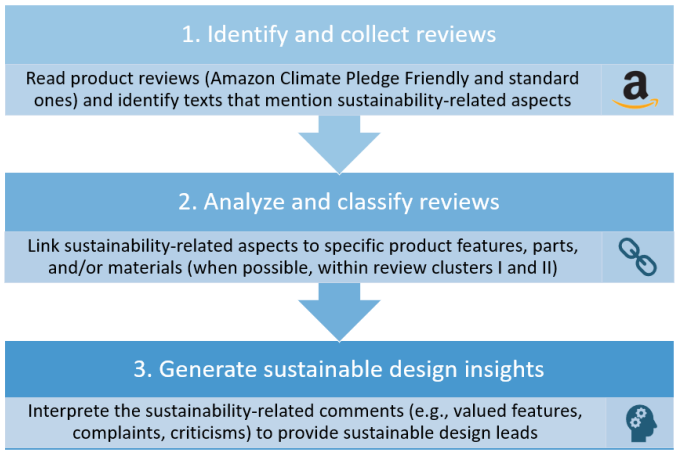

**FIGURE 2:** SYNOPSIS OF THE RESEARCH, ANALYSIS AND INTERPRETATION PROCESS

In the first step, to be well aligned with the purpose of this study (i.e., design for sustainability), the focus is made on identifying consumer – technical or electronics – products having at least one Amazon Climate Pledge Friendly certification and a sufficient number of reviews (at least 20 reviews for each rating range, i.e., from 1 to 5 stars, see justification hereafter). Amazon Climate Pledge Friendly products correspond to products having one or more sustainability certifications among the 19 ecolabels available on Amazon [13]. Then, a standard or conventional counterpart, with similar characteristics, in terms of features and price, is also identified for comparison purposes and to potentially generate additional insights.

In the second step, the authors analyze the top 20 reviews (by relevance and from verified purchases) for each rating range. In all, 100 individual customer reviews are manually annotated and reviewed for each product. This ensures that a sufficient number of reviews is processed for interpretation (in the next step) while still being feasible in terms of time and efforts for the authors to do so on several products. When available, sustainability-related aspects are highlighted, then the reviews are individually and manually classified within one of the three following clusters: (i) reviews in Cluster I state purposefully/intentionally sustainability-related aspects and/or performance, (ii) reviews in Cluster II mention aspects or features of the product that can be indirectly associated to its sustainability performance, and (iii) reviews in Cluster III contain no sustainability-related text at all. The general classification scheme in three sustainability-related clusters is depicted in Figure 3.

In the third step, reviews from clusters I and II – i.e., those including sustainability-related aspects – are interpreted by the authors to come up with sustainable design leads for the products considered. Notably, comparing the reviews between a product with environmental certification and its conventional alternative, additional insights are provided, and statistical analyses are made. El Dehaibi et al. (2019) provided a (non-exhaustive) list of topics to look for in reviews for each sustainability aspect [2]: (i) social aspects include health and safety, education, communication support, human rights; (ii) environmental aspects include material use, energy and water consumption, product durability, air and water emissions, waste and recycling; and, (iii) economic aspects include product price, cost-saving marketing, profit, business growth, and job creation. Out of the 100 reviews manually examined per product, the number of reviews that mention sustainability-related aspects is revealed and discussed. If insightful design suggestions can be extracted from the reviews, this process can be automated using machine learning tools and natural language processing techniques.

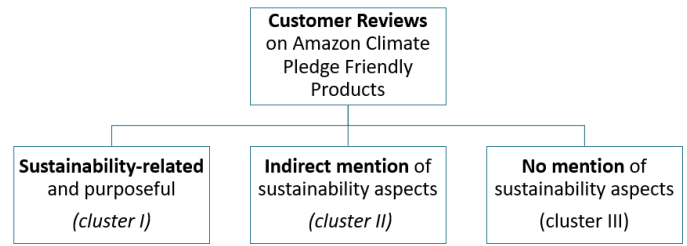

**FIGURE 3:** CLASSIFICATION MODEL OF PRODUCT REVIEWS IN THREE SUSTAINABILITY RELATED CLUSTERS

### 2.2 Products selection and description
For the case studies, the choice is oriented toward a comparison between two products within a similar price range and having the same overall functionality, one being certified with the Amazon Climate Pledge Friendly label and one being a basic counterpart without environmental claims. As the Climate Pledge Friendly program identifies products that improve at least one criterion relating to sustainability, it can be expected that the reviews on these products have a higher probability to mention sustainability-related aspects. In fact, ecolabels aim to foster the move toward more environmentally-friendly consumption patterns, as well as to increase the environmental standards of the products [14]. Moreover, the awareness and information that consumers get on the environmental footprint of a product through an ecolabel can stimulate pro-environmental behaviors or thoughts [3, 15].

For reference, according to the ISO categorization, there are three types of environmental "claims" for products [4]: (i) Type I being voluntary environmental labels based on a multi-criteria system which analyses the entire product life cycle, subject to external certification by an independent body [ISO 14024]; Type II being environmental self-declarations (such as "Recyclable", "Bio-based", "Compostable") by manufacturers, or product distributors without the intervention of an independent certification board [ISO 14021]; and (iii) Type III, being ecolabels which include declarations based on pre-established parameters which contain a quantification of the environmental impacts associated with the product life cycle. They should be subject to an independent check and presented in a transparent and comparable format [ISO 14025]. Amazon Climate Pledge

Friendly products belong to this third category. In practice, once an Amazon Climate Pledge Friendly product is identified as suitable for this study, a conventional alternative is selected within the "list of products related to this item" suggested by the Amazon web page, ensuring similar features, price range, and technical details.

Among the six Amazon Climate Pledge Friendly product categories – namely apparel, beauty, computers & office, electronics, grocery, health & household – the focus is put on technical and engineering products (i.e., from the computers and electronics categories) that have more complexity for product developers to actually design for sustainability. Within the category "computer & office", 755 results were found for Climate Pledge Friendly products (as of March 1$^{st}$, 2021), including laptops, printers, monitors, from a variety of brands such as HP, ASUS, BenQ. Within the category "electronics", 477 results were found for Climate Pledge Friendly products, including USB cables, printers, or smartphones.

**TABLE 1:** DETAILS AND STATISTICS ON TWO AMAZON CLIMATE PLEDGE FRIENDLY PRODUCTS (AS OF 3/1/2021)

| Product | HP Chromebook 14-inch Laptop | HP OfficeJet Pro 8025 Printer |
|---|---|---|
| Certification | EPEAT | EPEAT |
| Sustainability-related information (in product description) | "Environmentally conscious: Low halogen, mercury-free display backlights, arsenic-free display glass Energy Star and EPEAT certified." | "Sustainable design: This inkjet printer is made from recycled plastics and other electronics up to 15% by weight of plastic." |
| Average rating | 4.6 out of 5 | 4.4 out of 5 |
| Global ratings (#) | 3,375 | 9,923 |
| 5-star reviews (%) | 76.3% | 71.5% |
| 4-star reviews (%) | 13.5% | 10.1% |
| 3-star reviews (%) | 4.2% | 3.1% |
| 2-star reviews (%) | 1.6% | 2.1% |
| 1-star reviews (%) | 4.4% | 7.2% |

In all, three complementary Amazon Climate Pledge Friendly products were selected, namely: (i) a laptop, (ii) a printer, both having the Electronic Product Environmental Assessment Tool (EPEAT) certification (EPEAT products are assessed against criteria including energy use and have a reduced sustainability impact across their lifecycle), and (iii) a USB cable having the Compact by Design certification (Compact by Design products have less air, water, or packaging per use, which reduces the carbon footprint of shipping). Descriptive statistics for the two EPEAT products are shown in Table 1, including the sustainability-related features in the product description, the number of global ratings, the average ratings, and the distribution of reviews among the 5-star rating scale. In the next section, reviews associated with these products, as well as those from their respective counterparts, are further examined, classified, and interpreted to generate sustainable design leads.

### 3. RESULTS AND INTERPRETATION

### 3.1 Insights from sustainability-related reviews

In this sub-section, through the manual reading and processing of 100 reviews for each of the products mentioned in sub-section 2.2., the reviews are classified within the three clusters as shown in Table 2, and examples of reviews from clusters I and II are given and analyzed.

**TABLE 2:** DISTRIBUTION OF THE NUMBER OF REVIEWS MENTIONING SUSTAINABILITY-RELATED ASPECTS

| Product | HP Chromebook 14-inch Laptop | | | HP OfficeJet Pro 8025 Printer | | |
|---|---|---|---|---|---|---|
| Review cluster | I | II | III | I | II | III |
| 5-star reviews (/20) | 3 | 4 | 13 | 1 | 5 | 14 |
| 4-star reviews (/20) | 1 | 3 | 16 | 0 | 3 | 17 |
| 3-star reviews (/20) | 0 | 0 | 20 | 1 | 1 | 18 |
| 2-star reviews (/20) | 0 | 1 | 19 | 0 | 2 | 18 |
| 1-star reviews (/20) | 0 | 0 | 20 | 0 | 2 | 18 |

In Figure 4, extracts of two reviews with mentions of sustainability-related aspects for an Amazon Climate Pledge Friendly laptop are provided. The text highlighted in the top review indicates sustainability-related features that are directly valued by the customer (cluster I). The text highlighted in the bottom review indicates sentiments or criticisms that can be related to the sustainability performance of the product (cluster II). Other sustainability-related aspects mentioned by customers include the following, within the 5-, 4-, and 2-star reviews of this product: "Great laptop it's not a super thick laptop but who wants big thick laptop lol but overall built great battery life amazing."; "These are light and the battery lasts a lot longer than expected."; "Refurbished but better than new!"; "Lightweight and good battery."; "It is thin and lightweight."; "Lightweight, long battery use without the need of charging.".

Similarly, a sample of two reviews mentioning sustainability-related features for an EPEAT printer is given in Figure 5. Additional sustainability-related comments include, among the 5-star reviews: "The HP OfficeJet Pro 8025 is smartly designed and made of high-quality materials."; "Not only is its footprint smaller, its faster too!"; "Finally, the printer plastic feels great, but the plastic doors tend to bend easily on slight pressure.". Among the 4-star reviews: "[…] But, it prints sharp, is fast and you can replace the ink cartridges separately, so if you only run out of one color, that's all you need to replace."; "The cost to replace the non-functioning printer head was almost the same price as this whole unit. While this printer is smaller and lighter it is also more poorly constructed. It feels cheap and for the price, is cheaply made."; "For the price of $119 it is okay for a starter printer, but not for serious printing. I would send it back, but have destroyed the packaging. I will instead recycle it and go for a more robust printer.". Among the 2-star reviews: "Printer magically eats ink."; "Printers are designed to waste ink, this one is worse than usual.". Among the 1-star reviews: "This printer is cheaply made; the paper tray is flimsy and has 2 little plastic prongs to shift the paper upwards.".

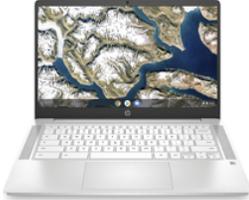

**FIGURE 4:** SAMPLE OF REVIEWS ON AN AMAZON CLIMATE PLEDGE FRIENDLY LAPTOP WITH EPEAT CERTIFICATION

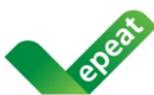

**FIGURE 5:** SAMPLE OF REVIEWS ON AN AMAZON CLIMATE PLEDGE FRIENDLY PRINTER WITH EPEAT CERTIFICATION

These consumer reviews are further analyzed in sub-section 3.3 to develop eco-improvement for these products based on feedback addressing sustainability-related features. But first, in the next sub-section, additional reviews made on similar products but without being environmental certified are investigated to provide further insights. Also, a supplementary and closest comparison between two cables, one having the Compact by Design Amazon certification and one without environmental certification, is conducted.

**3.2 Comparison between eco and non-eco products**

In the list of products related to the Amazon Climate Pledge Friendly laptop reviewed in sub-section 3.1, a laptop sharing similar specifications (in terms of performance, screen size, weight and price) and ratings, but without the environmental certification has been selected. The purpose here is to investigate if complementary (similar or different) sustainability-related comments could be found in the first top 100 reviews. This laptop is the Lenovo Chromebook, and the reviews mentioning sustainability-related aspects are the following: "Very light, solid feel to the keyboard."; "Solid. Reliable. Battery life is great, can go for hours at the library without needing to be plugged in and I'm using a higher brightness. Fast, lightweight, and cost-effective."; "Some suggestions for improvement: better screen resolution and contrast adjustment. And the power cord should be pin rather than a mini - plug . I'm concerned that the interior connection may get loosened over time as the computer gets moved about....bumping the cord will unseat that interior connection and possibly diminish performance."; "The battery stays charged well and it's lightweight."

In the same vein, a printer having comparable features to the Amazon Climate Pledge Friendly one has been chosen for comparison purposes. This printer is the Lexmark B3442dw, which does not have any environmental certification on the Amazon page, but claims some sustainability-related features in the product description as follows: "Durable: Steel frame and long-life imaging components mean it's built to last", and "Sustainable: Two-sided printing is standard, and built-in energy-saving modes." Key sustainability-related reviews include the following: "It is a nice size (not too large or too small) and it's sturdy steel frame makes it feel much more durable and substantial when compared to plastic versions."; "The ink lasts forever. Maybe more expensive but I will save hundreds each year on ink."; "Overall build quality of this printer is really solid, like most Lexmark printers that I've used recently. Some other manufacturers have gone for lightweight designs which are easy to move around but just don't feel nearly as sturdy. Lexmark is using nice thick plastics where plastic is used and plenty of metal in the structure."; "Prints both sides to save paper".

In the next sub-section, these sustainability-related reviews are combined, for each respective product category, in order to yield eco-improvement that are valued by customers. But first, let us compare an additional product (cable) having another Amazon Climate Pledge Friendly certification with its counterpart having no environmental certification.

The characteristics of the two cables being compared are given in Table 3. Note that the USB C to HDMI Cable designed and manufactured by Uni has the "Compact by Design" certification. Compact by Design is a sustainability certification created by Amazon to identify products that, while they may not always look very different, have a more efficient design [13]. With the removal of excess air and water, these products require less packaging and allegedly become more efficient to ship. At scale, these slight differences in product size and weight may lead to significant carbon emission reductions. Interestingly, both products have a non-negligible number of reviews that state sustainability-related aspects, as synthesized in Table 4.

**TABLE 3:** COMPARISON (SPECIFICATIONS AND REVIEWS) BETWEEN THE TWO CABLES (AS OF 3/1/2021)

| Product | USB C to HDMI 6ft Cable, by Uni | USB C to HDMI 6ft Cable, by QGeeM |
|---|---|---|
| Price | $17.99 | $15.99 |
| Certification | Compact by Design | None |
| Sustainability-related information (in product description) | "Sturdy and long-lasting design: braided nylon cable for extra durability, premium aluminum alloy casing for better heat dissipation." | No environmental certification, nor mention of sustainability-related aspects in the product description. |
| Average rating | 4.7 out of 5 | 4.5 out of 5 |
| Global ratings (#) | 33,456 | 9,420 |
| 5-star reviews (%) | 80.3% | 75.3% |
| 4-star reviews (%) | 10.1% | 11.2% |
| 3-star reviews (%) | 3.7% | 4.8% |
| 2-star reviews (%) | 1.5% | 2.1% |
| 1-star reviews (%) | 4.4% | 6.6% |

**TABLE 4:** DISTRIBUTION OF THE NUMBER OF REVIEWS FOR THE TWO CABLES

| Product | USB C to HDMI 6ft Cable, by Uni | | | USB C to HDMI 6ft Cable, by QGeeM | | |
|---|---|---|---|---|---|---|
| Review cluster | I | II | III | I | II | III |
| 5-star reviews (/20) | 4 | 3 | 13 | 5 | 3 | 12 |
| 4-star reviews (/20) | 1 | 3 | 16 | 2 | 4 | 14 |
| 3-star reviews (/20) | 1 | 3 | 16 | 2 | 2 | 16 |
| 2-star reviews (/20) | 1 | 1 | 18 | 0 | 0 | 20 |
| 1-star reviews (/20) | 0 | 0 | 20 | 0 | 2 | 18 |

For each cable, a sample of reviews is available in Figures 6 and 7. Highlights in the top review indicate sustainability-related features that are directly valued by the customer (cluster I). Highlights in the bottom indicate features and/or complaints that can be related to the sustainability performance of the product (cluster II).

For the Amazon Climate Pledge Friendly cable, here are additional extracts of sustainability-related comments made by consumers. Among the 5-star reviews: "Both USB-C end and HDMI end are made of aluminum material, looks and feels good quality. The black and grey braided nylon cable part is pretty solid, seems like it's going to last for a long time."; "I just started using this cord two days ago, so I'm not ready to review durability just yet. However, the materials used for this cord are excellent."; "The product came beautifully packaged and has all the major regulatory certifications. They have a lifetime warranty and are very proactive with supporting their customers (although I haven't needed their support yet). The ends of the cable are covered in high-quality aluminum that feels great in your hands, and the rope cable is extremely flexible. They got the details right."; "Solidly made cable- braided nylon sheath with rugged connections"; "The cable is covered with their advertised braided nylon, which I feel adds to the quality of the product. It seems more durable because of it.". Among the 4-star reviews: "The quality of this cable is top notch. It's light but strong with braided casing, and connectors are well designed. It is a little on the stiff side, but that comes from the tough shielding."; "One potential drawback is that, while the cable seems pretty high quality, there aren't flex points on the cord, so right near the HDMI port the cable can quickly go like 90° bent without some kind of mediating material - not sure it that matters to most, but I need as much minimal space behind my monitor, as I prop the laptop behind the monitor for the camera to function."; "It has a braided cord, gives it additional protection and is very sturdy.". Among the 2-star reviews: "Since the cable is braided, it's hard to stow away, Uni needs to add a cable fastening re-usable ties."; The material quality of the product is very durable and has superior performance until now.".

Additionally, here are the key sustainability-related reviews on the standard cable (i.e., without the environmental certification). Among the 5-star reviews: "The cord is durable thanks to the double braiding. The USB-C and HDMI ends come with protectors, so that when you're storing it away, no dust or other particles can damage them."; "I honestly cannot attest to the longevity or durability of the braided cable at this early point but it appears to be of high quality and very well made. I am impressed with the lifetime warranty."; "The braided cable has a good solid feel"; "From a durability perspective, I'm not expecting any issues due to the build quality, but of course durability is something that proves itself out over time so if there are any anomalies or negatives I will be sure to stop back and update this review.". Among the 3-star reviews: "Purchased because I needed something durable and long. The length was perfect. Unfortunately, it stopped working after about a month. I was disappointed.". Among the 1-star review: "I used this cable to connect to my monitor for about a year, but it has slowly declined in quality."

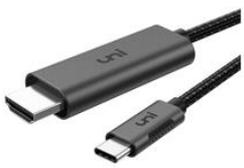

**FIGURE 6:** SAMPLE OF REVIEWS ON AN AMAZON CLIMATE PLEDGE FRIENDLY CABLE, CERTIFIED COMPACT BY DESIGN

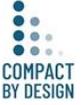

**FIGURE 7:** SAMPLE OF REVIEWS ON A CONVENTIONAL CABLE, WITHOUT ENVIRONMENTAL CERTIFICATION

### 3.3 Potential sustainable design leads

Based on the extraction, analysis, and human interpretation (here, the authors) of sustainability-related comments (either valued features or complaints), sustainable design improvements – for the three product types investigated in sub-sections 3.1 and 3.2 – that can be induced or deduced from these reviews are proposed, as summarized in table 5, and discussed. Note that the actual footprint and the potential impact savings associated with these solutions need to be adequately assessed by life cycle assessment or another environmental evaluation method.

For the cables, the braided nylon solution can be considered as a commendable sustainable design practice valued by customers – for durability at an affordable price – in comparison to simple plastic cords. Another suggested improvement to potentially enhance the durability of the cables, which can be subjected to frequent or repetitive bending constraints, is to design "flex points" on the cord, to bend the cable without altering the joints between the cord and the cable ends. As one mentions that the cable extremities could be hot and damage surrounding equipment inputs (such as a TV input) in the long run, a material choice valued by customers is premium aluminum alloy casing for better heat dissipation. Note that even the lowest-rated reviews (1- or 2-star reviews) were praising the durability of the cables, but complaining about technical aspects such as connectivity or degraded image quality.

For the printers, first, as several reviews mentioned the low robustness for sensitive or key parts made of plastic, a metallic (steel or aluminum) frame (e.g., for paper tray) seems a commendable sustainable alternative. Note that, here, trade-offs between the lightweight designs appreciated by most customers and the robustness of heavier solutions have to be taken into account. Second, a low consumption of inks and papers is highly valued by consumers. As such, the "prints both sides" feature appears to be a must-have sustainable solution. Lastly, as one mentioned the difficulty and cost to replace just one part of the printer compared to the price of buying a new one, modular printers could be a more sustainable alternative valued as well by customers. Other complaints were not related to sustainability-related features but were mostly addressing several issues related to the installation and the printing speed.

For the laptops (chromebook type), a great duration and lifetime of the battery is a key sustainable feature. Then, for such laptops, customers seem to appreciate the thinness and lightweight of the design, while one mentioned a potential fragility due to these environmentally-friendly features. Therefore, maintaining the desired thinness and lightweight levels while adding some reinforcement parts is a valuable direction for the design of more durable and sustainable laptops.

**TABLE 5:** SUSTAINABLE DESIGN INSIGHTS AND ASPECTS VALUED FROM CUSTOMERS, BASED ON ONLINE REVIEWS

| Products | Commendable sustainable design solutions, valued by customers |
|---|---|
| Cables | Braided nylon material for the cord |
| | Protective caps on connector ends |
| | Design of "flex points" on the cord |
| | Aluminum alloy casing for heat dissipation |
| Printers | Plastic for lightweight with metallic frame for robustness and durability on key mobile parts |
| | Double-sided printing capability |
| | Design for modularity (easy parts replacement) |
| Laptops | Thin and lightweight |
| | Reinforcement parts |

Overall, it can be stated that: (i) a segment of consumers – for both product categories, i.e., with or without the environmental certification – is showing an interest in the sustainability performance of the products they bought; (ii) they mention the sustainability-related features and aspects they value or would like to have through their online reviews; (iii) it is possible to interpret and translate these reviews in sustainable design leads or solutions for product designers.

Note that the interpretation of sustainable design insights suggested in this paper has been done manually on the basis of (only) hundreds of reviews analyzed, among the thousands of reviews often available for these kinds of products. As a consequence, it can be envisioned that further sustainable design leads valued by customers could be generated if one is able to analyze and process more reviews, e.g., through automation using artificial intelligence, including machine learning tools, and natural language processing techniques/algorithms. This perspective is further discussed in the conclusion section, as well as in another paper submitted to this conference [8].

One can imagine additional sustainable design insights that could be induced from online product reviews. For example, in situations where the majority of customers are complaining about the same product defects (e.g., leading to an early failure of a specific part), this could help designers fine-tune the product to make it more durable. Also, diving into the reviews of environmentally-certified products that are allegedly more sustainable, one can discover an unexpected and non-sustainable usage pattern by a group of customers that, in the end, makes the product less environmentally friendly (e.g., incorrect management of a battery, inappropriate maintenance). By identifying and being aware of these cases, the designers could modify the product to avoid such situations, e.g., by designing a kind of "Poka-Yoke" to ensure sustainable use.

In some other cases, customer feedback could help product developers to know more about how the final users are dealing with the product not only during its use phase but also at its end-of-life [16]. For instance, to prevent products from being landfilled at an early stage, some researchers analyzed customer reviews telling functionality problems and repair solutions [17], to improve the design and durability of their products accordingly. These analyses have been undertaken with the aim to find the top components that cause frequent failure, as well as to know the main causes of failure, and *in fine* to gain knowledge on how to fix a given failure based on the customer intelligence and experiences.

To close this Results section, here are additional takeaways we learned from this empirical study. First, consumers can perceive, by their own, sustainability-related aspects that are not mentioned by the manufacturer in the product description. Second, instead of mentioning directly the environmental repercussion (such as carbon reduction, or mitigation of resource depletion) of a particular feature, consumers are more likely to mention sustainability-related aspects that are directly beneficial for them (such as energy and consumables savings, or an increased durability enabled by a robust design or specific material choice).

## 4. CONCLUSION AND PERSPECTIVES

While online product reviews are increasingly seen as a valuable resource for product designers and developers, the analysis and exploitation of online product reviews mentioning sustainability-related aspects have the potential to help design more sustainable products that are still under-exploited. In the present study, by analyzing manually the top 100 reviews on two products of the same category (one with the Amazon Climate Pledge Friendly certification, and one without), and this for three different product types (laptop, printer, cable), it has been observed that a non-negligible number of reviews (around 15% on average based on the products and reviews analyzed here) mention relevant sustainability-related aspects. These reviews can include sustainable product characteristics valued by the customers, or complaints that can be related to a low durability or sustainability performance of a given product. Several examples have been given and discussed in the Results section.

On this basis, if we are able to automate this process – (i) collect (and filter out fake and meaningless) online product reviews; (ii) identify reviews addressing sustainability-related aspects; (iii) interpret the reviews to provide product developers with sustainable design improvement directions – further sustainable design insights could be mined and discovered out of the thousands of reviews available for many products. Note also that by diversifying the review sources, e.g., considering expert reviews from specialized platforms in complement to the consumer reviews available on Amazon or people commenting about new products on social media [18], further insights could be elicited.

For the future of sustainable design, it is indeed key to identify opportunities to integrate sustainability data and sustainable design methods into digitalized products and design processes [12]. The recent progress on artificial intelligence, machine learning, and natural language processing methods, is one promising opportunity, but not without challenges [8]. According to Isaksson and Eckert [5], product designers will increasingly be empowered by advancements in artificial intelligence capabilities to design the desired behavior before defining the system structure. Data analytics can actually help discover usage patterns and drive adaptations of products as well as feedback to product developers [5].

In fact, such methods have a great potential to solve problems where human processing capabilities are insufficient, as well as to expand the design space possibilities [19, 20]. For instance, designers could use, adapt or develop new natural language processing algorithms to support more automatic and intelligent knowledge extraction for sustainable design. Concretely, deploying existing natural language processing techniques, such as topic modeling and sentiment analysis, it is possible to extract the main product features that customers are commenting – praising, or complaining – about. Then, by building an *ad hoc* natural language processing dictionary and well-trained machine learning model, it could be possible to identify sustainability-related features. Next, using or developing a suitable aspect-based sentiment analysis algorithm, it could be possible to know to what extent the customers value these features, and *in fine* hone in a given product accordingly.

This empirical research work allows to open up on promising perspectives and timely research questions for future work, as well as to provide initial elements to answer them, such as "how product sustainability is perceived and valued by consumers?", or "do environmental labels on products posted by manufacturer drive customers to talk more about sustainability-related aspects in reviews?". According to recent studies [4], "the majority of consumers stated that, among products that cost the same, they would choose the ones with a lower impact on the environment, and ecolabels are one of the most important signals to the market to help this process".

Yet, one has to bear in mind that customers' inclination for sustainability attributes cannot guarantee their veracity. As illustrated in Figure 8, it becomes essential to further study and quantify the potential correlations or mismatches between: (i) the environmental claims made by manufacturers or retailers (e.g., in the product description or through an ecolabel), (ii) the perception [21] of sustainable features by consumers (e.g., captured through online product reviews), and (iii) the real impact of products on the environment (e.g., quantified through life cycle assessment [22]).

In future work, the insights obtained from the reviews could be backed up by a more robust LCA approach to determine which features are more "sustainable" than others (e.g., lightweight or durable, heavier materials), and thus to define the "ground truth" for what is considered a sustainable design recommendation, and what is not. Actually, a common understanding between developers and customers on sustainability values and features is critical for sustainable production and consumption. In this direction, Kwok et al. (2020) noticed a gap between how product sustainability information is perceived by customers and how this impacts the purchasing behavior of customers [23]. Through a specific case study, the authors started discussing the potential of (i) identifying prioritized sustainability attributes using sustainability design space, and (ii) applying machine learning to model customer preferences.

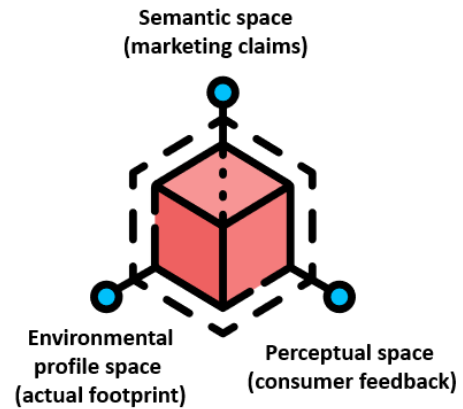

**FIGURE 8:** MAPPING OF DIFFERENT SUSTAINABILITY-RELATED SPACES TO FURTHER INVESTIGATE AND LINK